\documentclass[prb,twocolumn,amsmath,amssymb,superscriptaddress,citeautoscript]{revtex4}

\usepackage{epsfig}
\usepackage{graphicx}
\usepackage{wasysym}

\usepackage{amsmath}
\usepackage{amssymb}
\usepackage{epsfig}
\usepackage{graphicx}
\usepackage{wasysym}
\usepackage{bbm}

\newcommand \be{\begin{equation}}
\newcommand \ee{\end{equation}}
\newcommand \bes{\begin{equation*}} 
\newcommand \ees{\end{equation*}}
\newcommand \bea{\begin{eqnarray}}
\newcommand \eea{\end{eqnarray}}
\newcommand \beas{\begin{eqnarray*}} 
\newcommand \eeas{\end{eqnarray*}}
\newcommand \bfg{\begin{figure}}
\newcommand \efg{\end{figure}}
\newcommand \bfgs{\begin{figure*}} 
\newcommand \efgs{\end{figure*}}
\newcommand \bwt{\begin{widetext}}
\newcommand \ewt{\end{widetext}}

\newcommand \nd{{\vphantom{\dagger}}} 
\newcommand \bra{\langle}
\newcommand \ket{\rangle}
\newcommand \fig[1]{FIG.~\ref{#1}}
\newcommand \eqtn[1]{(\ref{#1})}
\newcommand \tbl[1]{TABLE~\ref{#1}}

\newcommand \vecs{\vec{s}}
\newcommand \vect{\vec{\tau}}

\newcommand \vecr{\vec{r}}
\newcommand \veck{\vec{k}}
\newcommand \ex{\hat{x}}
\newcommand \ey{\hat{y}}
\newcommand \im{\mathbbm{i}} 

\newcommand \dif{{\rm d}} 

\newcommand \idmat{\mathbbm{1}} 
\newcommand \mirror{m} 

\newcommand \schwinger{a}

\begin{document}
\title{A Z$_2$ spin-orbital liquid state in
 the square lattice Kugel-Khomskii model}
\author{Fa Wang}
\author{Ashvin Vishwanath}
\affiliation{Department of Physics, University of California,
 Berkeley, CA 94720}
\affiliation{Materials Sciences Division,
 Lawrence Berkeley National Laboratory, Berkeley, CA 94720}

\date{Printed \today}

\begin{abstract}
We argue for the existence of a liquid ground state in a class of
square lattice models of orbitally degenerate insulators. Starting
with the SU(4) symmetric Kugel-Khomskii model, we utilize a Majorana
Fermion representation of spin-orbital operators to access novel
phases. Variational wave functions of candidate liquid phases are
thus obtained, whose properties are evaluated using variational
Monte Carlo. These states are disordered, and are found to have
excellent energetics and ground state overlap ($>40\%$) when
compared with exact diagonalization on 16 site clusters. We conclude
that these are spin-orbital liquid ground states with emergent nodal
fermions and Z$_2$ gauge fields. Connections to spin 3/2 cold atom
systems and properties in the absence of SU(4) symmetry are briefly
discussed.
\end{abstract}

\maketitle
\section{Introduction}
In correlated insulators, the degrees of freedom that remain at low
energies are spin and orbital degeneracy. At low temperatures one
usually obtains an ordered state described essentially by a
classical variable, the Landau order parameter. Ground states that
are not described by the Landau framework are expected to possess
strikingly new properties. While they are known to occur in one
dimensional systems,
an important question is whether they arise in bulk two and three
dimensional materials. Theoretical studies have largely focused on
quantum spin systems. While model Hamiltonians for spin liquids
exist, one needs special conditions like strong frustration to
ensure that the spins do not order. Otherwise, even for spin 1/2
quantum fluctuations are not typically strong enough to destroy
order.  On the other hand, orbital degeneracy in insulators can
enhance quantum fluctuations\cite{Zaanen,Khallulin,BosscheZhangMila},
destroying order and possibly lead to a spin-orbital liquid state. An
experimental illustration is provided by the insulating spinels
MnSc$_2$S$_4$ and FeSc$_2$S$_4$. The former, a pure S=5/2 system,
magnetically orders below $2$Kelvin. The latter, a spin S=2 system,
is identical in most respects except that it involves a twofold
orbital degeneracy. In contrast to the spin system, it is found to
remain spin disordered down to the lowest temperatures of
$30$mKelvin \cite{Loidl1}.

Here, we theoretically study a simple spin 1/2 square lattice model
with the minimal twofold orbital degeneracy. Such a spin-orbital
Hilbert space is realized for e.g. with the
$d_{3z^2-r^2},\,d_{x^2-y^2}$ orbitals of Ag$^{2+}$, Cu$^{2+}$ or (low
spin) Ni$^{3+}$ in an octahedral environment. We focus on a model
that captures the effect of enhanced quantum fluctuations from
orbital degeneracy, and argue for a liquid ground state in this
case. We give theoretical arguments for the stability of this phase
as well as well as a numerical Monte Carlo study of a variational
wave function. The latter is found to have extremely good energetics
for our Hamiltonian, and allows us to characterize this phase beyond
the simple fact that it is disordered. The low energy collective
excitations of the liquid state are captured by an emergent Z$_2$
gauge field, coupled to Dirac like fermionic excitations with
fractional spin and orbital quantum numbers. The ground state
wavefunction is strongly entangled, and can be thought of as a
product of three Slater determinant wave-functions.

Realistic spin-orbital Hamiltonians tend to be rather complicated
with several exchange couplings that are strongly direction
dependent. Moreover, a linear coupling between the orbital degrees
of freedom and the Jahn-Teller phonons can quench coherent orbital
dynamics. However, for sufficiently strong exchange interactions,
the coupling to phonons can be ignored, and the orbitals can be
taken to be quantum degrees of freedom. Since we are interested here
in the general effects of enhanced quantum fluctuations from orbital
degeneracy, we follow \cite{BosscheZhangMila} and others in
considering a model that treats all four states on a site
symmetrically, i.e. the SU(4) symmetric spin orbital model. This
will allow for comparisons and is a useful starting point. We show
later that our essential conclusions are unchanged on perturbing
away from this high symmetry point.

The high symmetry SU(4) model may, in fact, have a direct physical
realization. We point out that a model of e$_g$ orbitals on certain
high symmetry lattices, like the diamond lattice, can have spin
orbital Hamiltonian that are nearly SU(4) symmetric.  A different
setting for this physics has been opened up by the recent
experimental developments on the trapping and cooling of alkaline
earth atoms. Confining these to the sites of an optical lattice
leads to SU(N) symmetric magnetic models.  The nuclear spin here
provides the $N$ flavors, and, given the weak dependence of
scattering lengths on nuclear spin, leads to SU(N) symmetric
exchange interactions, which, for fermionic atoms, will be
antiferromagnetic\cite{Hermele}. Fermionic alkaline-earth like atoms
of $^{173}$Yb have been cooled to quantum degeneracy
\cite{Fukuhara1}, while the Mott state of the bosonic $^{174}$Yb has
been recently realized \cite{Fukuhara2}. A different realization may
be provided by spin 3/2 cold atoms, such as $^{132}$Cs confined to
the sites of an optical lattice. At unit filling, one has four
states per site, and although the physical symmetry is only that of
spin rotations, the small difference in scattering lengths imply
only a weak breaking of SU(4) symmetry. It was pointed out in
\cite{Wu} that even including these differences only breaks
the symmetry down to SO(5)$\times$Z$_2$ in the low energy limit.

\section{The Model:} We study the SU(4) symmetric Kugel-Khomskii
model \cite{KK,Li} on the {\em square lattice}:
\begin{equation}
  H=\frac{J}{4}\sum_{<ij>}(\vecs_i\cdot\vecs_j+1)(\vect_i\cdot\vect_j+1)
 \label{Eqn:Model}
\end{equation}
 where $\vecs$ are the spin-1/2 Pauli matrices and
 $\vect$ are the Pauli matrices acting on the two
degenerate orbital states. We consider the antiferromagnetic(AFM) case
($J>0$) and set $J=1$ hereafter. The high symmetry of this model
implies that the three spin operators $\vecs$, three orbital
operators $\vect$ and nine spin orbital operators $\sigma^a\tau^b$
all appear with equal weight. These are fifteen generators of SU(4).
It should be noted that symmetry does not uniquely define a model,
one also needs to specify the representation of the symmetry group
appearing at each site. Here the fundamental representation appears
and an SU(4) singlet can only be formed between four sites.

The model (\ref{Eqn:Model}) was numerically studied in \cite{BosscheZhangMila}
using exact diagonalization
(the model suffers from a sign problem in the spin-orbital basis) on system
sizes of up to 4$\times$4.
The ground state ($E=-17.4$)
is an SU(4) singlets at zero wavevector. Simple trial states, such
as with spin-orbital order , or a box singlet state, have much
higher average energy ($E=0$ and $E=-12$ respectively) pointing to
the importance of quantum fluctuations. This motivates the study of
spin orbital liquids as candidate ground states.

The idea of resonating valence bonds \cite{PWA,RKS}, provides an
intuitive picture of the quantum spin liquid. A more formal approach
that is easier to generalize is the slave particle formalism, where
the spin is decomposed into `partons', e.g.
 $\vec{s}_r = \sum_{\sigma,\,\sigma'}
 \schwinger^\dagger_{\sigma r}\vec{\sigma}\schwinger_{\sigma' r}^\nd $,
where the $\vec{\sigma}$ are the Pauli matrices, and
$(a^\dagger_{\uparrow r},\, a^\dagger_{\downarrow r})$ creates a
boson (Schwinger boson) or fermion with spin $(\uparrow, \downarrow)$
at site $r$, and the constraint
 $\sum_{\sigma}\schwinger^\dagger_{\sigma r}\schwinger_{\sigma r} = 1$
is imposed at every site.
One then makes a mean field decomposition to obtain a quadratic Hamiltonian,
and the constraint is then imposed by projecting the wave function to obtain
a variational ground state. The state so obtained is a candidate
spin-liquid wave function. This procedure can be generalized in a
straightforward way to the spin-orbital Hilbert space at hand, by
introducing a four component $\schwinger_{\sigma}$ and the constraint above
at every site \cite{ShenZhang,Mishra}.
The physical spin and orbital operators are again bilinears of
 $\schwinger_\sigma$.
However, there are some drawbacks to this straightforward generalization.
The bosonic parton representation cannot treat the SU(4) symmetric
point,
while the fermionic parton theories necessarily lead to Fermi
surface states which can be hard to stabilize as ground states.

Below, we will sidestep these difficulties by showing that this
problem admits a third, physically distinct, parton representation
in terms of Majorana Fermions. This representation, which has not
previously been applied to two dimensional systems, offers us many
advantages. Besides being more economical (in terms of expanding the
Hilbert space in the minimal fashion), it leads to liquid states
with a $Z_2$ gauge group, whose low energy physics is well
understood and know to exist as stable phases. We emphasize here
that the projected wavefunction obtained from this Majorana parton
representation of the SU(4) model is distinct from the 'Schwinger'
fermion representation, involving four fermionic $a_\sigma$
operators.

\section{Majorana Parton Formulation:} We first point out the group
isomorphism SU(4)$\equiv$SO(6). The 15 generators of the latter are
dimension six antisymmetric real matrices $L_{\mu\nu}^A$, where
$A=1\dots 15$ and $\mu,\,\nu=1\dots 6$. An operator representation
of this algebra is obtained by introducing six Majorana fermions
$(\chi_1,\,\dots,\, \chi_6)$ which satisfy the anticommutation
relations $\{\chi_\mu,\,\chi_\nu \}=2\delta_{\mu\nu}$. The operators
 $\hat{O}^A = \frac14 L^A_{\mu\nu}\chi_\mu\chi_\nu$,
where summation over repeated indices is assumed, reproduce the commutation
relations for SO(6) generators. The Majorana Fermions transform as
SO(6) vectors.

We now use the group isomorphism to obtain a representation of the
spin-orbital operators, in terms of Majorana fermions. It is helpful
to write the set of Majorana fermions as a pair of three component vectors
 $(\vec{\theta}_r,\,\vec{\eta}_r)$ where, e.g.
 $\vec{\theta}_r=(\theta_{1r},\,\theta_{2r},\,\,\theta_{3r})$ and we
have introduced site indices $r$. The spin and orbital operators can
then be written in the compact form:
\be
 \vec{s}_r = -\frac{\im}{2}\vec{\eta}_r \times \vec{\eta}_r,\quad
 \vec{\tau}_r = -\frac{\im}{2} \vec{\theta}_r \times \vec{\theta}_r
\label{Eqn:spinRep} \ee
 and $s^{\mu}_r \tau^{\nu}_r = -{\im}\eta_{\mu r} \theta_{\nu r}$,
 which automatically obey the expected algebra. Note, the sign of
the Majorana fermion operators can be changed without affecting the
physical operators. This Z$_2$ redundancy is connected to the fact
that the Hilbert space is now enlarged - since each Majorana fermion
corresponds to $\sqrt{2}$ degrees of freedom, we have
$(\sqrt{2})^6=8$ states whereas there are only $4$ physical states
per site. The excess states can be removed by implementing a Z$_2$
constraint at each site: first define the operator $\nu_r$ commutes
with the physical operators (which are fermion bilinears) and is
idempotent $\nu_r^2=1$.
\begin{equation}
 \begin{split}
  \nu_r & \equiv \im \theta_{1r}\theta_{2r}\theta_{3r}
\eta_{1r}\eta_{2r}\eta_{3r}\\
  \nu_r & = 1, \, \forall r
 \end{split}
 \label{Eqn:constraint}
\end{equation}
Hence implementing the constraint in (\ref{Eqn:constraint}),
restricts us to the physical Hilbert space. This operator generates
the $Z_2$ gauge transformation on the Majorana fermions
 $\theta_{\alpha r}\rightarrow -\theta_{\alpha r},\,
 \eta_{\alpha r}\rightarrow -\eta_{\alpha  r} $.

The model in (\ref{Eqn:Model}) can be written in these variables as:
\begin{equation}
 H=\sum_{<jk>}\left [1-(1/8)\left (\im\vec{\eta}_{j}\cdot\vec{\eta}_{k}
 +\im\vec{\theta}_{j}\cdot\vec{\theta}_{k}\right )^2\right ]
\end{equation}
When supplemented by the constraint (\ref{Eqn:constraint}), this is
an exact rewriting of the model. Here, the $SO(6)$ symmetry of the
model is explicit. The quartic nature of the Hamiltonian requires an
approximation. We begin with a mean field treatment and use it to
generate variational states in which the constraint is treated
exactly.

%


In the context of models with only spin 1/2 (and no orbital degrees of freedom) we note that a representation utilizing three Majorana fermions per site,
 where the spin operator $\vec{s}_r$ is given by an expression
identical to that in Eqn.~\ref{Eqn:spinRep}, has been
studied \cite{Tsvelik}. However, we point out that this is distinct from our current formalism since a single site constraint cannot be applied to generate the physical Hilbert space.
Nevertheless, this provides an alternate parton approach - for example, with an even number of sites,
one can view half of the spins as `orbital pseudo-spins',
then the spin problem will be artificially converted to a spin-orbital problem,
(although without SU(4) symmetry).
Then the gauge fixing, and construction of complex fermions and
variational wave functions can be proceeded in the same fashion as
in this paper.
But this artificial discrimination of spin and `orbital pseudo-spin'
usually will superficially break lattice symmetry.
Another way to construct a Hilbert space for the spin 1/2 only model, is to introduce a
fourth Majorana fermion on every site, and the set of four fermions
satisfies a product constraint as in Eqn.~\ref{Eqn:constraint} \cite{Kitaev}.
This has the benefit of being formulated with a unique single site constraint.
However, unfortunately it turns out that these four fermions are just the real and imaginary
parts of the two Schwinger fermion operators, and do not lead to a
new representation. The constraint is the familiar one of requiring
single occupancy of the Schwinger fermions.


{\bf Mean Field Theory and Gutzwiller Projection:} With real mean
field parameters $\chi_{jk}$ ($=-\chi_{kj}$),  we have:
\be
 H_{\rm MF}=\sum_{<jk>}\left [1-(\im \chi_{jk}/4)
 \left (\vec{\eta}_{j}\cdot\vec{\eta}_{k}
 +\vec{\theta}_{j}\cdot\vec{\theta}_{k}\right )+\chi_{jk}^2/8\right ]
 \label{equ:Hmf1}
\ee
In self consistent mean field theory $\chi_{jk}= \im\langle
(\vec{\eta}_{j}\cdot \vec{\eta}_{k} +\vec{\theta}_{j}\cdot
\vec{\theta}_{k}) \rangle_{MF}$. For convenience we combine the 6
Majorana fermions into 3 complex fermions:
$c^\dagger_{\alpha r}=(1/2)(\eta^\nd_{\alpha r}+\im\theta^\nd_{\alpha r})$
which are more intuitive although the SO(6) symmetry is no longer explicit.
The constraint then is:
 $\sum_{a=1}^{3}c^\dagger_{\alpha r}c_{\alpha r}=0\ {\rm OR}\ 2$
(while the odd values of the site occupation are forbidden).
Writing $\vec{c}_r=(c_{1r},c_{2r},c_{3r})$, we have
 $\im(\vec{\eta}_{j}\cdot\vec{\eta}_{k}
  +\vec{\theta}_{j}\cdot\vec{\theta}_{k}) =
  2\im(\vec{c}_j^\dagger\cdot\vec{c}_k-\vec{c}_k^\dagger\cdot\vec{c}_j)$
i.e. the mean field theory simply involves fermions hopping with
pure imaginary amplitudes.   Such a band structure is automatically
particle-hole symmetric, which leads to half-filled bands for each
of the $c_{ra}$ fermions. Note, despite the imaginary hoppings the
mean field ansatz is time reversal symmetric if the hopping is
bipartite. The mean field wave function is simply a product of three
identical Slater determinants. While the specific Slater determinant
depends on the mean field ansatz, we make a few general observations
below. If we consider a system with $4N$ sites, required to obtain
an SU(4) singlet state, each Slater determinant $\Phi$ is a function
of $2N$ particle coordinates, corresponding to half filling.
Gutzwiller projecting the mean field state into the constrained
Hilbert space, yields a physical spin-orbital wave-function. In the
fermion representation, a site can either have no fermions (denoted
by $|0\rangle$), or two fermions, in which case there are three
states,
 $|X\rangle = c^\dagger_2 c^\dagger_3|0\rangle,\,
 |Y\rangle=c^\dagger_3 c^\dagger_1|0\rangle,\,
 |Z\rangle=c^\dagger_1 c^\dagger_2|0\rangle$.
These are related to
the spin orbital basis states via:
\begin{eqnarray*}
 |\sigma^z=\mp1,\tau^z=\pm 1 \rangle &=&(|0\rangle \pm \im|X
  \rangle)/\sqrt2\\
 |\sigma^z=\pm1,\tau^z=\pm 1\rangle &=&(|Y\rangle
  \pm \im|Z \rangle)/\sqrt2
\end{eqnarray*}
Given a configuration specified by the locations of the $|X\rangle$,
 $|Y\rangle$ and $|Z\rangle$ states (at sites $\{x_i$\},$\{y_j\}$ and
 $\{z_m\}$ respectively,
where $x_i,\,y_j,\,z_m$ are $3N$ distinct positions),
the spin-orbital wave function assigns an amplitude $\Psi\left [
\{x_i\},\{y_i\},\{z_i\} \right ]$ to it. Note, the locations of the
$|0 \rangle$ states are automatically specified. For an SU(4)
singlet we need equal numbers, $N$, of the four types of sites, so
$\{x_i\}=\{x_1,\dots,x_N\}$ etc. After the Gutzwiller projection we
obtain:
\begin{equation}
 \begin{split}
  &\Psi[\{x_i\},\{y_j\},\{z_m\}] \\
 = & \Phi[\{y_j \},\{z_m\}]\cdot\Phi[\{z_m\},\{x_i\}]\cdot\Phi[\{x_i\},\{y_j\}]
 \end{split}
 \label{Eqn:Gutzwiller}
\end{equation}
Thus the projected spin-orbital wave function is a product of three
Slater determinants with a lot of entanglement. We now apply this
formalism to specific models.

\section{One Dimensional Chain} This SU(4) symmetric nearest neighbor
model in 1D is very well understood and serves as a benchmark for
our technique. The only symmetric mean field ansatz is a uniform
$\chi_{r,r+1}=\chi$, leading to a dispersion
$\epsilon(k)=\chi\sin(k)$. We construct the resulting projected wave
function for $L$ site chain with $L=8,16\dots 128$ and antiperiodic
boundary conditions, {and evaluate its properties using
variational Monte Carlo}. The energy per site from the projected
wave functions extrapolated to the thermodynamic limit is $-0.8233$,
not far from the exact result\cite{Sutherland},
 $1-(1/2)\int_{0}^{1}\frac{x^{-3/4}-1}{1-x}\dif x=-0.8251$.
The leading term in the asymptotic spin correlation function is
$\cos(\pi r/2)/r^{1.5}$, consistent with theoretical and numerical
predictions \cite{Azaria,Affleck,Pati}. Note, these desireable
properties of the wavefunction only arise {\em after} projection.
{
(see \tbl{table:chain}).
}
\begin{table}
\caption{Results of the projected wave function for $L$-site chain
with antiperiodic boundary condition.
The second row shows energy per site for the projected wave function.
The third row shows the $L^{-1.5}$ scaling of
$s^z$-correlation functions[$s^z(x)$ is $s^z$ at position $x$].
}
\begin{tabular}{|r|rrrrr|}
\hline
$L$ & 8 & 16 & 32 & 48 & 64  \\
\hline
$\bra H \ket/L$ &
-0.8642 & -0.8332 & -0.8256 & -0.8242 & -0.8237 \\
\hline
$\bra s^z(0) s^z(L/2)\ket L^{1.5}$ &
1.66 & 1.82 & 1.90 & 1.92 & 1.96 \\
\hline\hline
$L$ & 80 & 96 & 112 & 128 & \\
\hline
$\bra H\ket/L$ & -0.8235 & -0.8234 &  -0.8233 &  -0.8233 & \\
\hline
$\bra s^z(0) s^z(L/2)\ket L^{1.5}$ &
1.98 & 1.94 & 1.97 & 1.97 & \\
\hline
\end{tabular}
\label{table:chain}
\end{table}

Interestingly, the $\pi/2$ wavevector of the dominant correlations
do not correspond to a natural wavevector of the mean field
dispersion, and arises entirely from projection. In contrast, this
wavevector is easier to understand on projecting a quarter filled
band, which arises in the standard fermionic representation of
spin-orbital operators $\schwinger_\sigma$. Remarkably, one can show
that the projected wave functions arising from this representation
and the Majorana fermion representation discussed above, are {\em
identical in one dimension}. We stress that this is a special
feature of one dimension, and in higher dimensions, the two will
lead to physically distinct states.
{
Details of the proof can be found in Appendix~\ref{app:proof}.
}


\section{Square Lattice:} The mean field states on the square lattice
 can be distinguished by the gauge invariant flux through the elementary
plaquettes e.g. $\chi_{jk}\chi_{k\ell}\chi_{\ell m}\chi_{mj}$ for the
plaquette $jk\ell m$. Translation and Time reversal symmetry
dictates that this flux must be uniform and can be either $0$ or
$\pi$. This leads to two distinct mean field states the uniform and
$\pi$ flux state ansatz.
The uniform states ansatz is
 $\chi_{\vecr,\vecr+\ex}=\chi_{\vecr,\vecr+\ey}=\chi$, where
 $\vecr=(x,y)$ is the position of lattice sites. The mean field
dispersion is $\epsilon(\veck)=\chi[\sin(k_x)+\sin(k_y)]$ for all
the three flavors, and has a square Fermi surface. However, the
uniform ansatz state has higher energy than the $\pi$-flux ansatz
both in mean field theory and
after projection, so we focus on the $\pi$-flux ansatz.

\begin{figure}
 \includegraphics{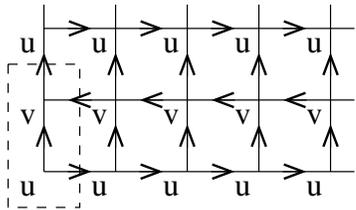}
 \caption{The $\pi$-flux ansatz on the square lattice.
  An arrow from site $j$ to $k$ means $\chi_{jk}>0$.
  Dashed lines enclose the doubled unit cell, with two sites $u$ and $v$.}
 \label{fig:piansatz}
\end{figure}

The $\pi$-flux ansatz is
$(-1)^y\chi_{\vecr,\vecr+\ex}=\chi_{\vecr,\vecr+\ey}=\chi$, as shown
in \fig{fig:piansatz}. The unit cell in mean field theory is
doubled, with $u$ and $v$ sublattices as shown.

After a Fourier transform, the mean field Hamiltonian
\eqtn{equ:Hmf1} is:
\begin{equation*}
 \begin{split}
 & H_{\rm MF}=(2+\chi^2/4)L^2\\
+&\chi\sum_{a,\veck}
  \begin{pmatrix}c_{\alpha,\veck,u}^\dagger & c_{\alpha,\veck,v}^\dagger
  \end{pmatrix}
  \begin{pmatrix}
   \sin k_x & \sin k_y \\
   \sin k_y & -\sin k_x
  \end{pmatrix}
  \begin{pmatrix}
   c_{\alpha, \veck,u} \\
   c_{\alpha, \veck,v}
  \end{pmatrix}
 \end{split}
\end{equation*}
where the sum over $\veck$ is over the $L^2/2$ (for $L\times L$ lattice)
$k$-points in the reduced($0\leq k_x < 2\pi,\, 0\leq k_y < \pi$)
Brillouin zone(BZ), and $a=\{1,2,3\}$. The above result
can be further diagonalized by a Bogoliubov transformation and
produce the two branches of the mean field dispersion
$\epsilon_{\pm}(\veck)=\pm\chi\sqrt{\sin^2 k_x + \sin^2 k_y}$. This
dispersion has two Dirac nodes at $\veck=(0,0)$ and $(\pi,0)$, with
isotropic dispersion in their vicinity. Including flavor indices, we
thus have 6 two-component Dirac fermions.

We use anti-periodic boundary conditions in both directions for
$L\times L$ lattices($L$ even) lattices. The $k$-points are then
$k_x=(2n+1)\pi/L,\ n=0\dots L-1;\ k_y=(2m+1)\pi/L,\ m=0\dots L/2-1$,
which avoids zero energy modes. Filling th negative energy states
gives us a Slater determinant mean field wave function for each of
the three fermion species. The Gutzwiller projected wave function is
then easily written down as (\ref{Eqn:Gutzwiller}), in terms of this
Slater determinant. Evaluating its properties however requires a
numerical variational (determinantal) Monte Carlo approach
\cite{Gros,Ceperley}. We generate a random initial basis state
having significant overlap with the mean field wave function. Random
pairs of sites are selected and updated with the Metropolis
rejection rule. 10000 `thermalization' sweeps($L^2$
pairwise updates) are performed before
measurements of physical quantities.
Measurements are done in 100000 sweeps.
The entire process is repeated 10 times to ensure
stability of results.

{\em Energetics:} The energy from the projected wave function is
listed in \tbl{table:Epersite} for $L$ up to 20. We notice that for
the $4\times 4$ lattice our total energy $-16.57J$ is close to the
ground state energy $-17.35J$ obtained in the exact diagonalization
study\cite{BosscheZhangMila}. More importantly, our energy lies
below the first excited state energy $-16J$ obtained in that study,
which already implies a {\em significant overlap} ($>42\%$) between
our wave function and the exact ground state wave function. Note, our
`variational' wave function has no variational parameter - which
makes this agreement more remarkable, especially given that there
are
24024 SU(4) singlet states already at this system
size\cite{BosscheZhangMila}.

\begin{table}
 \caption{Results of
  the projected wave function on a $L\times L$ square lattice with
  $\pi$-flux ansatz under anti-periodic boundary conditions in both directions.
  The second column shows energy per site.
 $Q_{\rm Box}$ in the third column and its relation to box order is defined
 in main text.
 The fourth column shows $L^{-4}$ scaling of spin correlation function when
 $L$ is not a multiple of four [$s^z(x,y)$ is $s^z$ at position $(x,y)$].
 Some entries are empty because the numerical errors are too large.
 }
\begin{tabular}{|r|lcr|}
\hline
$L$ & $\bra H\ket/L^2$ & $Q_{\rm Box}/L^2 $ & $-\bra s^z(0,0)s^z(L/2,0)\ket L^4$\\
\hline
 4 & -1.0357(4) & 2.0 & 11.14(2)  \\
 6 & -0.9238(3) & 1.7 & 20.46(5) \\
 8 & -0.9051(2) & 1.7 & 0.0(1)\\
10 & -0.8995(2) & 1.6 & 20.1(2) \\
12 & -0.8974(2) & 1.6 & 2.7(4)\\
14 & -0.8966(1) & 1.6 & 18.1(4) \\
16 & -0.8961(1) & 1.6 & \\
20 & -0.8956(1) & 1.6 & \\
24 & -0.8955(1) & & \\
\hline
 \end{tabular}
 \label{table:Epersite}
\end{table}

\begin{table}
\caption{Energy of several (variational) states for the SU(4) model on
 $4\times 4$ square lattice with periodic boundary condition.}
\label{table:Energetics}
\begin{tabular}{|l|r|}
\hline
state & energy($J$) \\
\hline
exact ground state\cite{BosscheZhangMila} & -17.35 \\
exact first excited state\cite{BosscheZhangMila} & -16 \\
projected SO(6) Majorana fermion mean field & -16.57 \\
projected SU(4) Schwinger fermion mean field & -6.38 \\
orbital ferromagnetic, spin AFM state\cite{Li,Gross} & -14.46\\
box(plaquette) ordered state\cite{Li} & -12 \\
four-sublattice SU(4) Neel state\cite{Li} & 0 \\
\hline
\end{tabular}
\end{table}
{ We have checked several other simple states on this $4\times 4$
lattice, they all show much higher energy, compared to the first
excited state in the exact spectrum. The comparison is shown in
\tbl{table:Energetics}. }

{\em Wavefunction properties:} We now provide evidence that the
resulting wave function is a spin-orbital liquid. First, we would
like to establish that it has no conventional order, to clearly show
it is not a conventional state. Next, the specific type of liquid
state being proposed - with emergent Dirac fermions and Z$_2$ gauge
fluxes - needs to be established.

\begin{figure}
 \caption{$s^z$ structure factor for the $\pi$-flux
  SO(6) projected wave function on a $20\times 20$ square lattice
  with anti-periodic boundary conditions.}
 \includegraphics{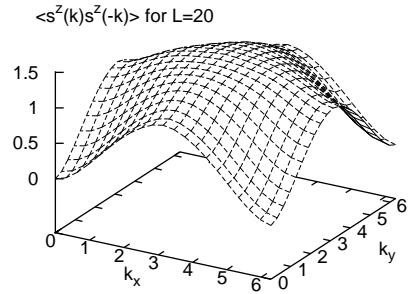}
 \label{fig:structurefactor}
\end{figure}

We first check for spin-orbital order. Given the SU(4) symmetry, it
is sufficient to compute the $s^z$ correlations which are found to
be rapidly decaying in space. The structure factor for various
lattice size was computed, e.g. \fig{fig:structurefactor} shows the
result for the $L=20$ lattice. A broad maximum at $(\pi,\pi)$ is
seen, but no Bragg-peak develops. We thus exclude
magnetic-orbital-ordering.

A more likely order is an SU(4) singlet
state that breaks lattice symmetry. This is the analog of the
Valence Bond Solid order for SU(2) magnets. However, SU(4) singlets
require at least four sites so box crystalline orders may arise. Two
such natural orders were proposed by Li, {\it et al.}\cite{Li}. In
both, the SU(4) singlets are formed on 1/4 of the elementary
plaquettes, but these are either arranged in a square lattice, or in
a body centered rectangular lattice with aspect ratio 2. Both box
orders have bond energies modulated at the wavevectors  $(\pi,0)$ or
$(\pi,\pi/2)$. We first check if our wave function has these
correlations by defining $E_{x,\veck}=\sum_{\vecr}e^{\im
\veck\cdot\vecr} (s^z \tau^z)_{\vecr}\cdot (s^z
\tau^z)_{\vecr+\ex}$. Then $Q_{\rm Box}=\bra
E_{x,\veck=(\pi,0)}^2\ket$ for a $L\times L$ lattice should scale as
$L^4$, if long range order is present. For example, for the perfect
square box state which is a product state of SU(4) singlets, $Q_{\rm
Box}=L^4/36+(13/9)L^2$. In the absence of order however, this
quantity will scale as $L^2$. For $L\le 20$ we did not observe $L^4$
scaling but rather good $L^2$ scaling, as shown in Column 3 of
\tbl{table:Epersite} indicating no sign or box order upto $400$ site
systems. An independent check is provided by modulating the mean
field parameters $\chi_{jk}$ to realize the box orders. The average
energy of the projected state is then compared against the
unmodulated wave function. We find that the energy always increases,
for both kinds of orders, pointing to the stability of the
unmodulated state. Within mean field theory alone, however, the
state is locally unstable to box modulation \cite{AffleckMarston}.
The more reliable projected energy study however point to the
opposite conclusion.

%
Motivated by a recent proposal of chiral SU(N) states in large-N
limit\cite{Hermele}, we also consider a chiral state
on the square lattice. We add to the $\pi$-flux state
pure imaginary hopping of fermions on the diagonal bonds
in such a way that each triangle has $+\pi/2$ flux.
This is a particular mass term of the Dirac fermions
and it opens a gap in the mean field dispersion, similar to the box order
mentioned above.
Indeed the mean field energy decreases with the diagonal hoppings.
However after projection the energy always increase after adding this term,
indicating stability against this chiral order. This distinction between
mean field and projected mean field energetics has been observed in
other projected wave function studies as well\cite{YingRan}.

\begin{figure}
\caption{$\tau^z$ structure factor for the $\pi$-flux SO(6)
projected wave function with average fermion filling $5/8$ for
one fermion specie(other two species are still half filled), on
 a $20\times 20$ square lattice with anti-periodic boundary conditions.}
\includegraphics{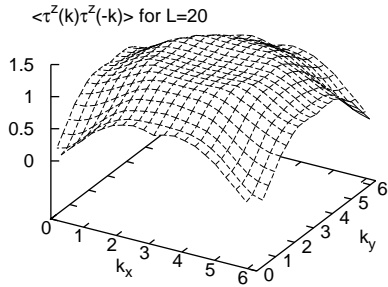}
\label{fig:ringcorrelation}
\end{figure}

\begin{figure}
\caption{(Color online) 
$\tau^z$ structure factor when $k_y=\pi$(a 1D cut
 of \fig{fig:structurefactor} and \fig{fig:ringcorrelation}) for
the $\pi$-flux SO(6)
projected wave function with average fermion filling
 $p=1/2$(green solid line with symbol $+$) and
 $p=5/8$(red dash line with symbol $\times$)
for one of the three species
on
 a $20\times 20$ square lattice with anti-periodic boundary conditions,
both curves are of arbitrary scale.}
\includegraphics{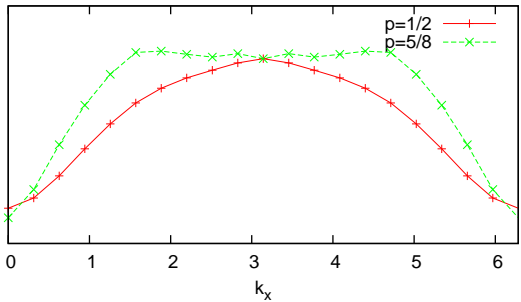}
\label{fig:ringcorrelationcut}
\end{figure}

We expect the spin-orbital liquid to be a nodal Z$_2$ state, i.e. it
contains emergent Z$_2$ gauge fields and nodal Dirac fermions that
behave like free particles at low energies. Establishing this
directly is more challenging - it is well known that observing the
Z$_2$ topological order of projected wave functions in the presence
of gapless gauge charged fermions is tricky \cite{IvanovSenthil} and
left to future work. Free nodal fermions would lead to spin and
orbital correlations that decay as $1/r^4$, which we check for by
computing $L^4 \bra s^z(0,0) s^z(L/2,0)\ket$ in a size $L$ system.
The fast decay limits us to $L \le 14$. The results are shown in
column 4 of \tbl{table:Epersite}. There are strong
commensuration effects which reduce the correlation when $L/2$ is an
even number. However, for the other three values of $L$ the
correlation seems to show the required scaling. Another indirect
evidence for such fermionization is the nature of these spin
correlations in the presence of a Zeeman field $\Delta H = -h\sum_r
\tau_r^z$, which leads to a shifted chemical potential for one fermion
specie. We find that the projected wave function now has a ring of
incommensurate correlations around $(\pi,\pi)$
(\fig{fig:ringcorrelation} and \fig{fig:ringcorrelationcut}).

{\bf Breaking SU(4) symmetry} It is natural to ask if the physical
conclusions derived above are stable when enlarged SU(4) symmetry of
our model is lost.
Since the gauge fluxes are gapped, a weak perturbation cannot lead
to confinement. Also, the gapless nodal fermions are actually
protected by discrete symmetries - one needs to break
lattice symmetry to gap the nodes,
as can be seen from an analysis of the fermion
bilinear terms(see Appendix~\ref{app:psg} for details).
The only physical difference that arises in the
lower symmetry case is that the chemical potential of the fermions
may not be at the nodal points. Hence the SU(4) symmetry is not
essential to our conclusions. In future\cite{FWAV}, we will apply
this analysis to realistic Hamiltonians with reduced symmetry and
search for liquid phases in those regimes.

\section{Physical Realizations}
{Since most natural spin orbit Hamiltonians are not near
the SU(4) symmetric point, it is natural to ask where we might
expect to find a model where the SU(4) symmetry is even
approximately realized. As discussed in the introduction, cold atom
systems in optical lattices provide some promising direction for
realization, if sufficient cooling of those magnetic Hamiltonians
can be achieved. In this section, we point out that even in solid
state systems, approximate SU(4) symmetry may be achieved, on
certain high symmetry lattices. In particular, we point out that on
the {\em diamond} lattice, if only nearest neighbor exchange is
considered, the interactions are close to the SU(4) point. Exchange
interaction arises from a combination of hopping and on site
interaction. The main observation is that due the high symmetry of
the diamond lattice, hopping matrix elements must be SU(4)
symmetric. The onsite interactions deviate from SU(4) symmetry due
to, for e.g., the Hunds interaction. However, these are typically a
fraction of the overall repulsion leading to nearly SU(4) symmetric
exchange.}

For a system of $d_{3z^2-r^2},\,d_{x^2-y^2}$ orbitals on the diamond
lattice with full lattice symmetry and without spin-orbital
coupling, we first prove that the electron hopping matrix elements
on nearest neighbor bonds have SU(4) symmetry. Denote the creation
operators of the two orbitals as
 $d_{1\alpha r}^\dagger$ and $d_{2\alpha r}^\dagger$ respectively,
where $\alpha $ is spin index, $r$ is site index.
The hopping amplitude on bond $<ij>$ is generically a $2\times 2$
matrix $t$, and the process is described by the term
 $\sum_{a,b,\alpha} d_{a\alpha i}^\dagger t_{ab}^\nd d_{b\alpha j}^\nd$.
Consider a bond from origin along the $(111)$ direction,
the reflection $x\to y,\,y\to x$ does not change this bond,
however the orbitals transform non-trivially
$d_a \to \sum_b(\sigma^z)_{ab} d_b$.
Since this reflection is a physical symmetry,
the electronic Hamiltonian should be invariant under its action,
thus we get $t=\sigma^z\cdot t\cdot \sigma^z$.
Similarly consider a threefold rotation
 $x\to y,\,y\to z,\,z\to x$,
it does not change the $(111)$ bond as well,
but the orbitals transform as
$d_a \to \sum_b[(-1/2)\sigma^z+\im (\sqrt{3}/2)\sigma^y]_{ab}d_b $.
Then we get
 $t=[(-1/2)\sigma^z-\im(\sqrt{3}/2)\sigma^y]\cdot t \cdot
 [(-1/2)\sigma^z+\im(\sqrt{3}/2)\sigma^y]$.
These two conditions on $t$ ensure that
 $t$ is proportional to identity matrix.
Thus we have proved that the hopping on $(111)$ direction preserves
both orbital and spin, namely is given by the term
 $t\sum_{a,\alpha}d_{a\alpha i}^\dagger d_{a\alpha j}^\nd$.
By lattice symmetry we conclude that all other nearest neighbor bonds
have this property. Therefore the nearest neighbor hoppings
have SU(4) symmetry. One should note that this proof cannot be extended to
next nearest neighbor and other generic hoppings.

However, the Coulomb interaction of these orbitals generically does not
have SU(4) symmetry.
The onsite Coulomb interaction is given by
the Kanamori parameters\cite{KanamoriParameters},
$U\sum_{a}n_{a\uparrow}n_{a\downarrow}
+U'\sum_{a < b}n_{a}n_{b}
+J\sum_{a < b,\alpha,\beta}
 d_{a\alpha}^\dagger d_{b\beta}^\dagger d_{a\beta}^\nd d_{b\alpha}^\nd
+J \sum_{a < b}(
 d_{a\uparrow}^\dagger d_{a\downarrow}^\dagger
 d_{b\downarrow}^\nd d_{b\uparrow}^\nd+{\rm h.c.})
$,
and approximately $U=U'+2J$.
The SU(4) symmetry is present only if $J=0$ and $U=U'$.
We usually expect that $J \ll U$, then SU(4) is an approximate symmetry
of the Hubbard model and thus an approximate symmetry of the derived
spin-orbital exchange model.

This type of systems may be realized in certain A-site spinels,
where the magnetic ions form a diamond lattice, and when only the
e$_g$ orbitals are active. One caveat is that the spinel structure
allows for the next nearest neighbor exchange strength to be fairly
large and even comparable to the nearest neighbor one, which may
significantly break the SU(4) symmetry. {Interestingly, the
experimentally discussed `spin-orbital' liquid candidate,
FeSc$_2$S$_4$, is an e$_g$ system on the diamond lattice
\cite{Loidl1}. However, it differs in two important respects from
the ideal model considered here. First, there is a magnetic moment
on each site that is Hunds coupled to the e$_g$ fermion, and second,
the further neighbor exchange interactions are believed to be
substantial in this material.}



We acknowledge support from NSF-DMR0645691, and discussions with M. Hermele.

\appendix
\section{Proof of the equivalence between
projected SO(6) Majorana mean field state and
projected SU(4) Schwinger fermion mean field state for 1D chain.}
\label{app:proof}

Consider a $4N$-site chain with periodic boundary condition.
The mean field wave function for the Majorana fermion
representation is
\be
\begin{split}
 |\Psi_{\rm MF}\ket=&
 \prod_{k=0}^{2N-1} \tilde{c}_{1,(2k+1)\pi/(4N)}^\dagger
 \prod_{k=0}^{2N-1} \tilde{c}_{2,(2k+1)\pi/(4N)}^\dagger \\
&\times
 \prod_{k=0}^{2N-1} \tilde{c}_{3,(2k+1)\pi/(4N)}^\dagger |0\ket
\end{split}
\ee
where $|0\ket$ is fermion vacuum, $\tilde{c}_{\alpha, k}$ is
the Fourier transform of real space fermion operator
 $\tilde{c}_{\alpha, k}=(4N)^{-1/2}\sum_{r} c_{\alpha, r} e^{-\im k r}$.
For a physically allowed real space configuration mentioned above
$ |\{x_i\},\{y_j\},\{z_m\}\ket
 =\prod_{i=1}^{N}( c_{2 x_i}^\dagger c_{3 x_i}^\dagger )
\prod_{j=1}^{N} ( c_{3 y_j}^\dagger c_{1 y_j}^\dagger )
\prod_{m=1}^{N} ( c_{1 z_m}^\dagger c_{2 z_m}^\dagger ) |0\ket
$,
the overlap with the mean field wave function is
\be
\begin{split}
& \Psi[\{x_i\},\{y_j\},\{z_m\}]
= \bra \{x_i\};\{y_j\};\{z_m\} | \Psi_{\rm MF} \ket \\
= & \Phi[\{y_j \},\{z_m\}]\cdot\Phi[\{z_m\},\{x_i\}]\cdot\Phi[\{x_i\},\{y_j\}]
\end{split}
\ee
where $\Phi$ is the $2N\times 2N$ Slater determinant for
 one fermion specie with the following matrix elements ($p,q=1,\dots,2N$)
\be
 \sqrt{4N}\cdot \Phi[\{x_i\},\{y_j\}]_{pq} =\left \{
 \begin{array}{ll}
   e^{\frac{\im \pi (2p-1) x_q}{4N}}, & q\leq N;\\
   e^{\frac{\im \pi (2p-1) y_{q-N}}{4N}}, & N < q.
 \end{array}\right.
\ee
Therefore the determinant is
\be
\begin{split}
 & \Phi[\{x_i\},\{y_j\}]
\\ = &
(4N)^{-N}\omega^{\cdot (\sum_{i}x_i+\sum_{j}y_j)}
 \prod_{i,j} (\omega^{2y_j}-\omega^{2x_i})
\\ & \times
 \prod_{i>i'} (\omega^{2x_i}-\omega^{2x_{i'}})
 \prod_{j>j'} (\omega^{2y_j}-\omega^{2y_{j'}})
\end{split}
\ee
where $\omega=e^{\im \pi/(4N)}$.
And the overlap is
\be
\begin{split}
& \Psi[\{x_i\},\{y_j\},\{z_m\}] \\
= &
(4N)^{-3N}\omega^{2\cdot(\sum_{i}x_i+\sum_{j}y_j+\sum_{m}z_m)}
\\ &\times
 \prod_{i,j} (\omega^{2y_j}-\omega^{2x_i})
 \prod_{j,m}(\omega^{2z_m}-\omega^{2y_j})
\\ &\times
 \prod_{m,i}(\omega^{2x_i}-\omega^{2z_m})
 \prod_{i>i'} (\omega^{2x_i}-\omega^{2x_{i'}})^2
\\ &\times
 \prod_{j>j'} (\omega^{2y_j}-\omega^{2y_{j'}})^2
 \prod_{m>m'} (\omega^{2z_m}-\omega^{2z_{m'}})^2
\end{split}
\ee

The mean field Hamiltonian for the standard `Schwinger' fermion representation
is
\be
\begin{split}
H_{\rm{MF},\schwinger} = & \chi \sum_{r}
 \sum_{\alpha=1}^{4} \schwinger^\dagger_{\alpha,r}\schwinger^\nd_{\alpha,r+1}
+{\rm h.c.}
-\mu\sum_{r}\sum_{\alpha=1}^{4}
 \schwinger^\dagger_{\alpha,r} \schwinger^\nd_{\alpha,r}
\\ = &
\chi\sum_{r}
 \sum_{\alpha=1}^{3} \schwinger^\dagger_{\alpha,r}\schwinger^\nd_{\alpha,r+1}
-\chi\sum_{r}
 {\schwinger'}^\dagger_{4,r}{\schwinger'}^\nd_{4,r+1}
+{\rm h.c.}
\\ &
-\mu\sum_{r}\sum_{\alpha=1}^{3}
 \schwinger^\dagger_{\alpha,r} \schwinger^\nd_{\alpha,r}
+\mu\sum_{r}
 {\schwinger'}^\dagger_{4,r}{\schwinger'}^\nd_{4,r}
\end{split}
\ee
where $\schwinger'$ is the particle-hole conjugate of the Schwinger fermion
 $\schwinger$.
The particle-hole transformation on the fourth specie is required for
the following projected wave function to represent a bosonic spin-orbital
wave function.
The quarter-filling mean field wave function for
the standard fermionic representation is
(we assume $N$ is even for simplicity)
\be
\begin{split}
 &|\Psi_{\rm MF}\ket_{\schwinger} \\
= &
 \prod_{k=-N/2}^{N/2-1}\tilde{\schwinger}_{1,(2k+1)\pi/(4N)}^\dagger
 \prod_{k=-N/2}^{N/2-1}\tilde{\schwinger}_{2,(2k+1)\pi/(4N)}^\dagger
\\ &\times
 \prod_{k=-N/2}^{N/2-1}\tilde{\schwinger}_{3,(2k+1)\pi/(4N)}^\dagger
 \prod_{k=-7N/2}^{-N/2-1}\tilde{\schwinger'}_{4,(2k+1)\pi/(4N)}^\dagger |0'\ket
\end{split}
\ee
where $\tilde{\schwinger}_{\sigma,k}$ is the Fourier transform of
the real space SU(4) `Schwinger fermions'
 $\tilde{\schwinger}_{\sigma,k}=(4N)^{-1/2}e^{-\im kr}\schwinger_{\sigma,r}$,
and $|0'\ket$ is the fermion `vacuum' that can be annihilated by
 $\schwinger_{1},\,\schwinger_{2},\,\schwinger_{3}$ and the
 particle-hole conjugate of the fourth specie $\schwinger'_{4}$.

A physically allowed real space configuration in this representation
is still labeled by three sets of $N$ distinct numbers
 $\{x_i\},\{y_j\},\{z_m\}$,
\be
\begin{split}
 |\{x_i\},\{y_j\},\{z_m\}\ket
 =  &
 \prod_{i=1}^{N}\schwinger_{1,x_i}^\dagger
 \prod_{j=1}^{N}\schwinger_{2,y_j}^\dagger
 \prod_{m=1}^{N}\schwinger_{3,z_m}^\dagger
\\ & \times
 \prod_{i=1}^{N}\schwinger_{4,x_i}^\dagger
 \prod_{j=1}^{N}\schwinger_{4,y_j}^\dagger
 \prod_{m=1}^{N}\schwinger_{4,z_m}^\dagger |0'\ket
\end{split}
\ee
The overlap of this configuration with the mean field wave function
is the product of four Slater determinants,
\be
\begin{split}
& \Psi_{\schwinger}[\{x_i\},\{y_j\},\{z_m\}]
= \bra \{x_i\},\{y_j\},\{z_m\}|\Psi_{\rm MF}\ket_{\schwinger}\\
= &
\Phi_{\schwinger}[\{x_i\}]\cdot\Phi_{\schwinger}[\{y_i\}]
\cdot\Phi_{\schwinger}[\{z_m\}]\cdot
 \Phi_{\schwinger'}[\{x_i\},\{y_j\},\{z_m\}]
\end{split}
\ee
The $N\times N$ Slater determinants $\Phi_{\schwinger}$ has
the following matrix elements($p,q=1,\dots,N$),
$
\Phi_{\schwinger}[\{x_i\}]_{pq}=(4N)^{-1/2}e^{\frac{\im \pi(2p-1-N)x_q}{4N}}
$.
The matrix elements of the $3N\times 3N$ Slater determinant
 $\Phi_{\schwinger'}$ is ($p,q=1,\dots,3N$)
\be
\begin{split}
& \Phi_{\schwinger'}[\{x_i\},\{y_j\},\{z_m\}]_{pq}
\\ = &
\left\{\begin{array}{ll}
 e^{\frac{\im \pi(2p-1-7N)x_q}{4N}},& q \leq N\\
 e^{\frac{\im \pi(2p-1-7N)y_{q-N}}{4N}},& N < q \leq 2N\\
 e^{\frac{\im \pi(2p-1-7N)z_{q-2N}}{4N}},& 2N < q \leq 3N
\end{array}\right.
\end{split}
\ee
Therefore the determinant of $\Phi_{\schwinger}$ is
\be
\Phi_{\schwinger}[\{x_i\}]=(4N)^{-N/2}
\omega^{(1-N)\cdot(\sum_{i}x_i)}
\prod_{i>i'}(\omega^{2x_i}-\omega^{2x_{i'}})
\ee
And the determinant of $\Phi_{\schwinger'}$ is
\be
\begin{split}
& \Phi_{\schwinger'}[\{x_i\},\{y_j\},\{z_m\}]
\\ = &
(4N)^{-3N/2}\omega^{(1-7N)(\sum_{i}x_i+\sum_{j}y_j+\sum_{m}z_m)}
\\ &\times
 \prod_{i,j} (\omega^{2y_j}-\omega^{2x_i})
 \prod_{j,m}(\omega^{2z_m}-\omega^{2y_j})
\\ &\times
 \prod_{m,i}(\omega^{2z_m}-\omega^{2x_i})
 \prod_{i>i'} (\omega^{2x_i}-\omega^{2x_{i'}})
\\ &\times
 \prod_{j>j'} (\omega^{2y_j}-\omega^{2y_{j'}})
 \prod_{m>m'} (\omega^{2z_m}-\omega^{2z_{m'}})
\end{split}
\ee
Finally the overlap between this basis state and
the mean field wave function is (note that $\omega^{8N}=1$)
\be
\begin{split}
& \Psi_{\schwinger}[\{x_i\},\{y_j\},\{z_m\}]
\\ = &
(4N)^{-3N} (-1)^{N^2}
 \omega^{(2-8N)\cdot(\sum_{i}x_i+\sum_{j}y_j+\sum_{m}z_m)}
\\ & \times
 \prod_{i,j}(\omega^{2y_j}-\omega^{2x_i})
 \prod_{j,m}(\omega^{2z_m}-\omega^{2y_j})
\\ & \times
 \prod_{m,i}(\omega^{2x_i}-\omega^{2z_m})
 \prod_{i>i'}(\omega^{2x_i}-\omega^{2x_{i'}})^2
\\ & \times
 \prod_{j>j'}(\omega^{2y_j}-\omega^{2y_{j'}})^2
 \prod_{m>m'}(\omega^{2z_m}-\omega^{2z_{m'}})^2
\\ = & (-1)^{N^2} \omega^{-8N(\sum_{i}x_i+\sum_{j}y_j+\sum_{m}z_m)}
\Psi[\{x_i\},\{y_j\},\{z_m\}]
\\ = & (-1)^{N^2} \Psi[\{x_i\},\{y_j\},\{z_m\}]
\end{split}
\ee Therefore these two projected wave functions for 1D chain are
identical. {The crucial property utilized was that the
Slater determinants appearing are Vandermonde determinants in one
dimension.} This property does not hold in higher dimensions.

\section{
Projective symmetry group analysis of
fermion bilinears in the
$\pi$-flux state on square lattice
}\label{app:psg}

For the $\pi$-flux ansatz in \fig{fig:piansatz} on an infinite lattice
the lattice group symmetries are realized as
follows(flavor index $\alpha$ omitted):
\beas
 T_x &:& c_{(x,y)}\to c_{(x+1,y)}\\
 T_y &:& c_{(x,y)}\to (-1)^x c_{(x,y+1)}\\
 R_{\pi/2} &:& c_{(x,y)}\to \frac{1+(-1)^y-(-1)^x+(-1)^{x+y}}{2} c_{(-y,x)}\\
 m_{y} &:& c_{(x,y)}\to (-1)^x c_{(-x,y)}
\eeas
Define a 4-compoent field
\bes
 \psi_{\alpha,\veck}=
 [c_{\alpha,\veck,u},c_{\alpha,\veck,v},
 c_{\alpha,(\pi,0)+\veck,v},c_{\alpha,(\pi,0)+\veck,u}]
\ees
linearize the dispersion around the Dirac point,
the low-energy Hamiltonian becomes
\bes
 \chi \sum_{\alpha,|\veck|\ll 1} \psi_{\alpha,\veck}^\dagger
 [k_x \idmat\otimes\mu^z + k_y\idmat\otimes\mu^x ]\psi_{\alpha,\veck}^\nd
\ees
where $\mu$ are Pauli matrices acting on the $u,v$ 2D space.
The original fermion in real space can be represented as
 $c_{\alpha,\vecr}=
  N_{\rm site}^{-1/2}\sum_{|\veck|\ll 1}e^{\im\veck\cdot\vecr}
 \{[1+(-1)^y]\cdot[(\psi_{\alpha,\veck})_1+(-1)^x (\psi_{\alpha,\veck})_4]
  +[1-(-1)^y]\cdot[(\psi_{\alpha,\veck})_2+(-1)^x (\psi_{\alpha,\veck})_3]\}/2$
where $\vecr=(x,y)$.
Then we have the transformation property of the $\psi$ field
\beas
 T_x &:& \psi_{(k_x,k_y)} \to \nu^z\otimes \idmat\psi_{(k_x,k_y)} \\
 T_y &:& \psi_{(k_x,k_y)} \to \nu^x\otimes \idmat\psi_{(k_x,k_y)} \\
 R_{\pi/2} &:& \psi_{(k_x,k_y)} \to (1/2)(1+\im\nu^y)\otimes(1+\im\mu^y)\psi_{(-k_y,k_x)} \\
 \mirror_y &:& \psi_{(k_x,k_y)} \to \nu^x\otimes\mu^x\psi_{(-k_x,k_y)} \\
\eeas
where $\nu^{x,y,z}$ are Pauli matrices acting on the 2D space of the two
Dirac nodes $(0,0)$ and $(\pi,0)$.
The low energy Hamiltonian is invariant under these transformations.
In the following we will prove that these symmetries prohibit mass term
and velocity anisotropy in the low energy theory.

Consider a general mass term $\psi^\dagger M \psi^\nd$ where $M$ is
a $4\times 4$ constant non-trivial(not proportional to identity)
Hermitian matrix.
$T_x$ and $T_y$ translation symmetries require that $M\propto\idmat\otimes\mu$
with $\mu$ be a $2\times 2$ constant Hermitian matrix.
$\mirror_y$ reflection symmetry requires $\mu\propto\mu^x$,
but this violate the $R_{\pi/2}$ rotation symmetry and is forbidden.

Consider a general velocity anisotropy term
 $\psi^\dagger(k_x M_1+k_y M_2)\psi^\nd$ with constant $4\times 4$ matrices
 $M_1$ and $M_2$.
Again $T_x$ and $T_y$ translation symmetries require that
 $M_{1(2)}\propto\idmat\otimes \mu_{1(2)}$.
$\mirror_y$ reflection symmetry requires $\mu^x\mu_1\mu^x=-\mu_1$ and
 $\mu^x\mu_2\mu^x=\mu_2$, therefore $\mu_2\propto \mu^x$.
Also consider 180$^\circ$ rotation $R_{\pi/2}^2$ symmetry
 $\psi_{(k_x,k_y)}\to -\nu^y\otimes\mu^y\psi_{(-k_x,-k_y)}$, it requires
$\mu^y \mu_1 \mu^y=-\mu_1$, thus $\mu_1\propto \mu^z$.
Now we have
 $\psi^\dagger (A k_x \idmat\otimes\mu^z + B k_y \idmat\otimes\mu^x)\psi^\nd$
with constants $A,B$. Use the 90$^\circ$ rotation symmetry $R_{\pi/2}$ we get
\bes
\begin{split}
 & A (-k_y) \idmat\otimes\mu^z + B k_x \idmat\otimes\mu^x
\\ = &
 A k_x \idmat\otimes[(1-\im\mu^y)\mu^z(1+\im\mu^y)]/2
\\ &
 +B k_y \idmat\otimes[(1-\im\mu^y)\mu^x(1+\im\mu^y)]/2
\\ = &
 A k_x \idmat\otimes\mu^x-B k_y \idmat\otimes\mu^z
\end{split}
\ees
Then we have $B=A$ and this term is just the kinetic energy term in the
Hamiltonian.

Consider a general bilinear term $\psi^\dagger M(k_x,k_y)\psi^\nd$ where
$M(k_x,k_y)$ is a $4\times 4$ hermitian matrix and a homogeneous polynomial
of $k_x,k_y$. From translation symmetries it must be of the form
 $\idmat\otimes [m_0(k_x,k_y)\idmat+m_1(k_x,k_y)\mu^x
  +m_2(k_x,k_y)\mu^y+m_3(k_x,k_y)\mu^z]$ where $m_{0,1,2,3}$ are
homogeneous functions of $k_x,k_y$ and are of the same order.
From $\mirror_y$ reflection symmetry,
\bes
 \begin{split}
 & m_0(-k_x,k_y)=m_0(k_x,k_y),\\
 & m_1(-k_x,k_y)=m_1(k_x,k_y),\\
 & m_2(-k_x,k_y)=-m_2(k_x,k_y),\\
 & m_3(-k_x,k_y)=-m_3(k_x,k_y)
 \end{split}
\ees
From $R_{\pi/2}$ rotation symmetry,
\bes
 \begin{split}
 & m_0(-k_y,k_x)=m_0(k_x,k_y),\\
 & m_3(-k_y,k_x)=-m_1(k_x,k_y),\\
 & m_2(-k_y,k_x)=m_2(k_x,k_y),\\
 & m_1(-k_x,k_y)=m_3(k_x,k_y)
 \end{split}
\ees
This requires that $m_0$ is of the form
 $\tilde{m}_0(k_x^2+k_y^2, k_x^2 k_y^2)$,
 $m_1$ is $k_y \tilde{m}_1(k_x^2,k_y^2)$,
 $m_2$ is $k_x k_y (k_x^2-k_y^2) \tilde{m}_2(k_x^2+k_y^2,k_x^2 k_y^2)$,
and $m_3$ is $k_x \tilde{m}_1(k_y^2,k_x^2)$,
where $\tilde{m}_0,\tilde{m}_1,\tilde{m}_2$ are arbitrary polynomials.


\begin{thebibliography}{}\label{sec:bibliography}

\bibitem{Zaanen} L. F. Feiner, A. M. Oles, and J. Zaanen, {Phys. Rev. Lett.} {\bf 78}, {2799} {(1997)}.
\bibitem{Khallulin} G. Khaliullin and S. Maekawa {Phys. Rev. Lett.} {\bf 85}, {3950} {(2000)}
\bibitem{BosscheZhangMila}Mathias van den Bossche, F.-C. Zhang, F. Mila, {Euro. Phys. J. B} {\bf 17}, {367} {(2000)}.

\bibitem{Loidl1} V. Fritsch, {\it et al.}, {Phys. Rev. Lett.} {\bf 92}, {116401} {(2004)}.

\bibitem{Hermele}M. Hermele, V. Gurarie, A. M. Rey, {arXiv:0906.3718}
\bibitem{Fukuhara1} T. Fukuhara, Yosuke Takasu, Mitsutaka Kumakura, and Yoshiro Takahashi,  {Phys. Rev. Lett.} {\bf 98}, {030401} {(2007)}.
\bibitem{Fukuhara2} T. Fukuhara, Seiji Sugawa, Masahito Sugimoto, Shintaro Taie, and Yoshiro Takahashi,  {Phys. Rev. A} {\bf 79}, {041604(R)} {(2009)}.
\bibitem{Wu}Congjun Wu, {Mod. Phys. Lett. B} {\bf 20}, {1707} {(2006)}.

\bibitem{KK}K. I. Kugel, and D. I. Khomskii, {Sov. Phys. Usp.} {\bf 25}, {231} {(1982)}.
\bibitem{Li}Y. Q. Li, Michael Ma, D. N. Shi, F. C. Zhang, {Phys. Rev. Lett.} {\bf 81}, {3527} {(1998)}.
\bibitem{PWA}P. W. Anderson, {Mater. Res. Bull.} {\bf 8}, {153} {(1973)}.
\bibitem{RKS}D. S. Rokhsar, and S. A. Kivelson, {Phys. Rev. Lett.} {\bf 61}, {2376} {(1988)}.
\bibitem{ShenZhang}S.-Q. Shen, G.-M. Zhang, {Europhys. Lett.} {\bf 57}, {274} {(2002)}.
\bibitem{Mishra}A. Mishra, M. Ma, and F.-C. Zhang, {Phys. Rev. B} {\bf 65}, {214411} {(2002)}.
\bibitem{Tsvelik} A. Tsvelik, {Quantum Field Theory in Condensed Matter
Physics}, {Cambridge University Press} {(1995)}.
\bibitem{Kitaev} A. Kitaev, {Annals of Physics} {\bf 321}, {2} {(2006)}.
\bibitem{Sutherland}B. Sutherland, {Phys. Rev. B} {\bf 12}, {3795} {(1975)}.
\bibitem{Azaria}P. Azaria, A. O. Gogolin, P. Lecheminant, A. A. Nersesyan, {Phys. Rev. Lett.} {\bf 83}, {624} {(1999)}.
\bibitem{Affleck}I. Affleck, {Nucl. Phys. B} {\bf 265}, {409} {(1986)}.
\bibitem{Pati}S. K. Pati, R. R. P. Singh, and D. I. Khomskii, {Phys. Rev. Lett.} {\bf 81}, {5406} {(1998)}.
\bibitem{Gros}C. Gros, {Annals of Physics} {\bf 189}, {53} {(1989)}.
\bibitem{Ceperley}D. Ceperley, G. V. Chester, and M. H. Kalos, {Phys. Rev. B} {\bf 16}, {3081} {(1977)}.
\bibitem{Gross}M. Gross, E. Sanchez-Velasco, and Eric Siggia {Phys. Rev. B} {\bf 39}, {2484} {(1989)}

\bibitem{AffleckMarston}I. Affleck, and J. B. Marston, {Phys. Rev. B} {\bf 37}, {3774} {(1988)}.
\bibitem{YingRan}Y. Ran, M. Hermele, P. A. Lee, X.-G. Wen, {Phys. Rev. Lett.} {\bf 98}, {117205} {(2007)}
\bibitem{IvanovSenthil}D. A. Ivanov, T. Senthil, {Phys. Rev. B} {\bf 66}, {115111} {(2002)}.
\bibitem{FWAV} Fa Wang and Ashvin Vishwanath, in progress.

\bibitem{KanamoriParameters} C. Castellani, C. R. Natoli, J. Ranninger, {Phys. Rev. B} {\bf 18}, {4945} {(1978)};
R. Fresard, G. Kotliar, {Phys. Rev. B} {\bf 56}, {12909} {(1997)}.

\end{thebibliography}
\end{document}